# A Mapping Study on Software Process Self-Assessment Methods

Thaísa C. Lacerda, Christiane G. von Wangenheim, Jean C.R. Hauck

Graduate Program in Computer Science - Department of Informatics and Statistics

Federal University of Santa Catarina (UFSC) - Brazil

thaisa.lacerda@posgrad.ufsc.br, c.wangenheim@ufsc.br, jean.hauck@ufsc.br

**Abstract:** Assessing processes is one of the best ways for an organization to start a software process improvement program. An alternative for organizations seeking for lighter assessments methods is to perform self-assessments, which can be carried out by an organization to assess its own process. In this context, the question that arises is which software process self-assessment methods exist and which kind of support they provide? To answer this question, a mapping study on software process self-assessment methods was performed. As result, a total of 33 methods were identified and analyzed, synthesizing information on their measurement framework, process reference model and assessment process. We observed that most self-assessment methods are based on consolidated models, such as CMMI or ISO/IEC 15504 with a trend to develop self-assessment methods specifically for SMEs. In general, they use simplified assessment processes, focusing on data collection and analysis. Most of the methods propose to collect data through questionnaires that are answered by managers or other team members related to the process being assessed. However, we noted a lack of information on how most of the assessment methods (AMs) have been developed and validated, which leaves their validity questionable. The results of our study may help practitioners, interested in conducting software process self-assessments, to choose a self-assessment method. This research is also relevant for researchers, as it provides a better understanding of the existing self-assessment methods and their strengths and weaknesses.

## 1. Introduction

Software process improvement is a validated way to increase product quality, to develop software with fewer resources in less time, to improve productivity increasing organizational flexibility and customer satisfaction and, thus, to allow software organizations to stand out from competitors (Harter et al. 2012; Kuilboer 2000; Boehm 2006; Paulk 1995; Solingen 2004; Kalinowski 2014). One of the best ways for an organization to start a software process improvement (SPI) program is to perform a process assessment to elicit the gap between its current practices and the ones indicated by a reference model 0(McCaffery et al. 2007).

In order to perform a process assessment, Process Assessment Models, also called Software Process Capability/Maturity Models (SPCMMs), are typically used (Gresse von Wangenheim et al. 2010). These models describe the life cycle processes and process generic practices (ISO/IEC 2012). There exist several SPCMMs, among the most widely used are the CMMI (SEI 2010) and ISO/IEC 15504[1] (2004) (Yucalar & Erdogan 2009). Yet, besides their popularity, they are mostly applied in large organizations, not becoming popular among small and medium enterprises (SME) and/or agile enterprises (Larrucea et al. 2016). This may be due to their detailed assessment procedure requiring considerable effort with significant costs, making their adoption often impossible for SMEs (Kar et al. 2012; Chang 2002; Yucalar & Erdogan 2009, McCaffery 2007, Gresse von Wangenheim et al. 2006; ISO IEC 29110-3; Abushama 2016). In addition, SMEs believe that software process assessments (SPAs) require a certain degree of detail that increases corporate bureaucracy (Kroeger 2014). Another reason that makes SPAs less attractive to SMEs is the difficulty of understanding and implementing them in practice (Yucalar & Erdogan 2009; Kalpana 2010). This fact leads many companies to seek more simplicity within the organizations' processes and as a result they are increasingly attracted to agile methods (McCaffery 2007). Despite these challenges that many organizations encounter to assess their processes, SPAs cannot be discarded as they allow identification of the organization's strengths and weaknesses and, thus, are a fundamental input to start a software improvement program (Habra 2008; Kar et al. 2012).

[1] Currently being replaced by the new standard ISO 33001.

Motivated by the need for less complex and more agile assessment methods (AMs), lighter assessment methods are developed, typically for SMEs or companies that use agile development methods, in the form of self-assessments. Self-assessments are the most common way to conduct a SPA in organizations that do not aim for certification (Patel and Ramachandran 2009). They are carried out by an organization to assess the capability/maturity of its own process, not requiring the involvement of external SPI experts. The sponsor of a self-assessment is normally internal to the organization, as are the member(s) of the assessment team responsible for collecting and analyzing data and reporting the assessment results (ISO/IEC 2004). Data collection is typically done by using a single method such as questionnaire, interview or workshop. Data analysis depends on the specific assessment purpose. For instance, to find gaps between the organization's current practices and the process assessment model (ISO/IEC 2004), to educate the organization on the requirements of a formal assessment method or to perform a benchmarking against other organizations (Blanchette et al. 2005).

The popularity of self-assessments is due to their low cost and accessibility (Patel and Ramachandran 2009; Abushama 2016). As self-assessments use the organization's own human resources and are less bureaucratic, they allow a more simplified way to perform a process assessment in a shorter period of time with fewer resources (Pino et al. 2010; Abushama 2016). In addition, in organizations where maturity is a new concept, self-assessments allow an easy way to improve the process by not significantly intruding on the daily routine of the organization (Blanchette et al. 2005). Self-assessment is also effective in generating an "ownership feeling" among managers, regarding the process quality, as it forces them to examine their own activities (Blanchette et al. 2005; Wiele et al. 2000).

Despite the benefits presented, self-assessments are not without shortcomings. Organizations using self-assessments find it difficult to plan the process and to allocate human recourses to lead and execute them. Another difficulty is the scarcity of literature regarding the "best" approach to perform a self-assessment as there is no guidance on which self-assessment method organizations should use (Ritchie and Dale 2000). This choice is especially important considering that different assessment methods may give different scores on attributes, not necessarily meaning that one method is better than the other, as sometimes two measures independently predict outcome criteria, each adding variance to the other (Ritchie and Dale 2000). This leads to the conclusion that applying a variety of data collection methods can be useful to construct a more comprehensive picture of the needs and gaps that need to be addressed (Goethals 2013).

Another issue is the possible absence of competent assessors. As assessors in self-assessment are not necessarily experts in process assessment and may not be familiar with SPCMMs, there is a considerable risk of misinterpretation of the process indicators (Blanchette et al. 2005). To mitigate this risk, data collection instruments used for the assessment must be explanatory in a way that non-experts may understand and adequately judge the indicators, e.g. preventing them from wrongly considering a Gantt chart to be a project plan. Furthermore, the response scale has also to be defined carefully, as assessors in self-assessment may not have sufficient experience to correctly classify the degree of satisfaction of an item on a finer grained scale (Saunders et al. 2009), e.g. deciding between partially achieved and largely achieved. Thus, if two or more points on a scale appear to have the same meaning, respondents may be puzzled about which one to select, leaving them to make an arbitrary choice (Krosnick et al. 2009). In order to minimize the assessment effort, data collection instruments should also be comprehensive enough to measure the essential information, but at the same time be succinct enough to encourage their completion.

Despite the extensive research on the development of lighter assessment methods and the recognition of the value of self-assessments, little systematic attention has been paid to this kind of assessment. Considering it an important element of process improvement programs, the question that arises is which software process self-assessment methods exist and what are their characteristics in terms of assessment process, techniques and stakeholders? Which process reference models and measurement frameworks do they use? Have these methods been developed and systematically evaluated?

Therefore, this research aims at answering these questions through the execution of a mapping study on software process self-assessment methods providing a broad review of the existing methods and categorizing them in terms of their measurement framework, process reference model and assessment process. We also analyze the methods with respect to their development and validation/evaluation. We expect the results of this work to be useful to practitioners to select and adopt self-assessments and, thereby, contribute to improve the quality of software products as well as to guide researchers on the improvement of these methods.

The remainder of this article is organized as follows: Section 2 presents the background on the subject addressed; Section 3 describes the research methodology; Section 4 and 5 present the definition and execution of the mapping study. Results of the data analysis are presented in Section 6 and the results are discussed in Section 7. Section 8 presents the conclusion.

# 2. Background

## 2.1. Process assessment elements

Process assessment is a disciplined assessment of the processes of an organization against a reference model compatible with a process assessment model (ISO/IEC 2004). ISO/IEC 15504 (2004) presents an assessment framework that defines the elements necessary to carry out a process assessment. Following ISO/IEC 15504 (2004), the framework for conducting assessments, called assessment method (AM), includes a process reference model, a measurement framework, an assessment model, as well as an assessment process. On the other hand, the reference model and assessment model are the same in CMMI (SEI 2011).

The process reference model describes a process life cycle, defining its purposes, process outputs and the relationships between them (ISO/IEC 2004). Typically, reference models are refined into activities (ISO/IEC 2012) or base practices (Coallier et al. 1994) that should be carried out, so the process might achieve its goal. Processes can be grouped into process areas (PAs), that are a group of related practices that when implemented satisfy important goals for making improvement in that area. Sometimes process areas are presented as dimensions or categories, which represent key elements of the process (SEI 2011).

The measurement framework provides a base for rating the capability of processes and/or the maturity of the organization, based on their achievement of defined process attributes. It typically includes three elements: process attributes, rating scale and a capability/maturity scale (ISO/IEC 2004). A process attribute represents measurable characteristics, which support the achievement of the process purpose and contribute to meeting the business goals of the organization (ISO/IEC 2004). A rating scale typically is an ordinal scale to measure the extent of the achievement of a process. The capability scale is composed by Capability Levels (CL) and represents the capability of the implemented process in increasing order, from not achieving the process purpose, to meeting current and projected business goals (SEI 2011). The maturity scale characterizes the maturity of the organization and each level builds on the maturity of the level below (Figure 1). Capability and maturity levels are typically represented using a staged or continuously scale system. The continuous representation uses Capability Levels to characterize capability relative to an individual Process Area (PA). A staged scale represents the Maturity Level (ML) of the organization's processes. Each maturity level is comprised of a set of process areas. To reach a certain maturity level the set of PAs must met a certain capability level (SEI 2011).

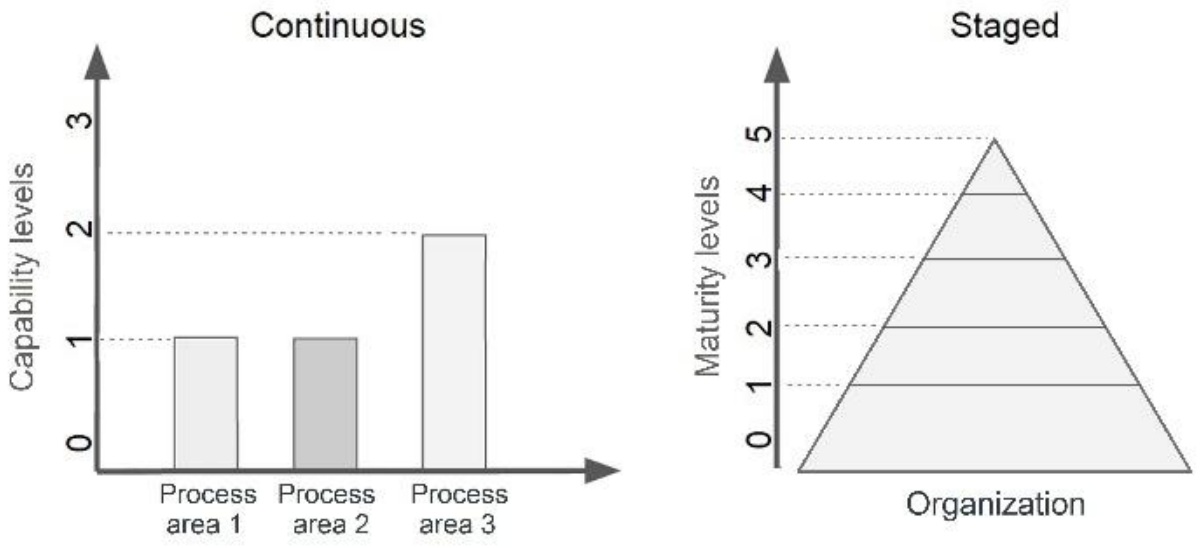

**Figure 1 Examples of continuous and staged scale (adapted from SEI (2010))**

The assessment process is a set of activities that must be performed to conduct an assessment (El Emam 1999). It contains activities such as planning, data collection, data validation, process attribute rating and reporting (ISO/IEC 2004), defining also their inputs and outputs (Figure 2). Each activity, in turn, can be performed by adopting specific techniques (such as interviews, workshops, meeting with stakeholders, presentations) and using specific tools (such as spreadsheets, templates or software systems). As part of the assessment process it also defines roles and responsibilities.

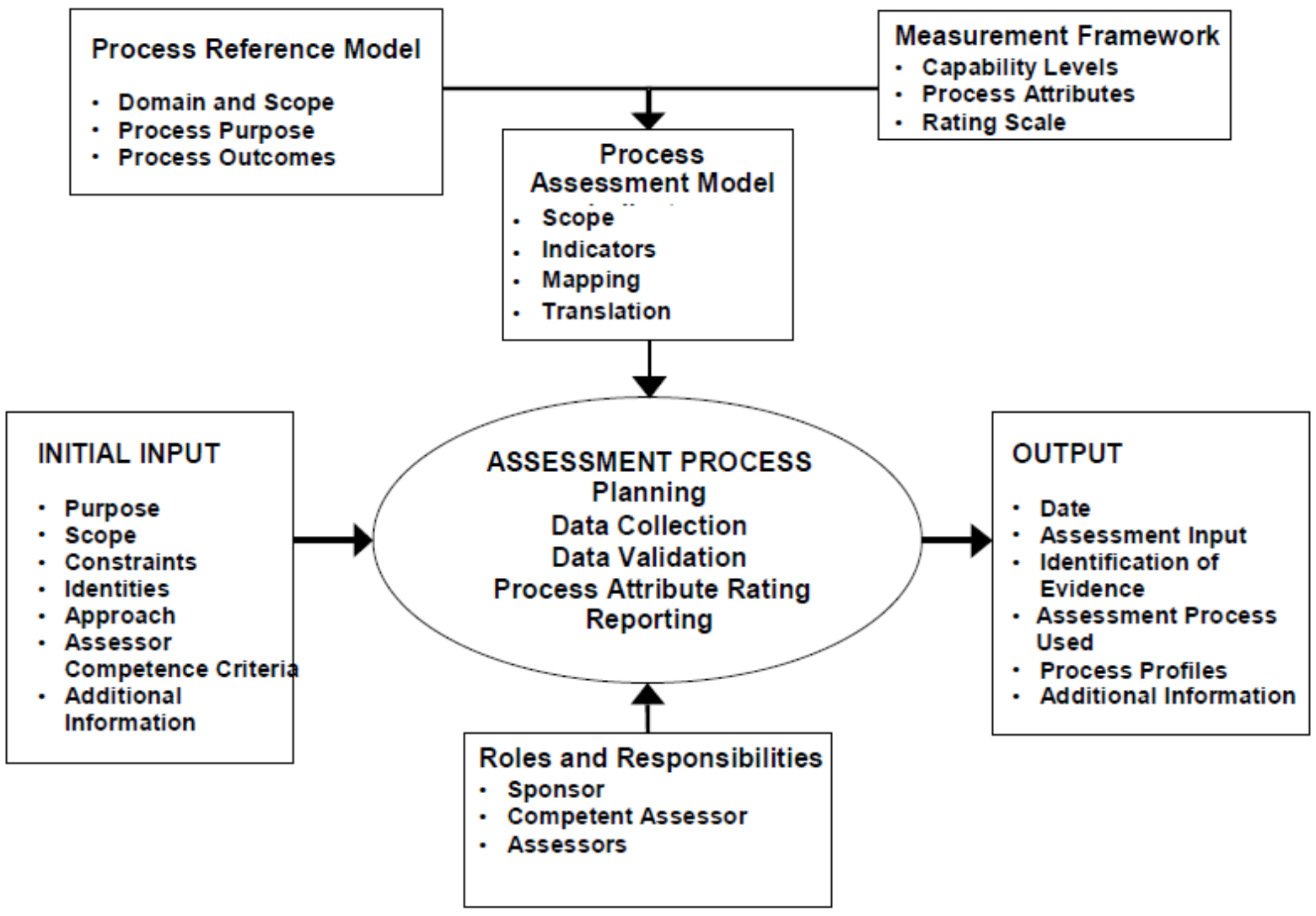

**Figure 2 Process assessment elements (ISO/IEC 2004)**

There are basically two ways for performing process assessments: as an independent assessment performed by a team external to the organization, or as a self-assessment performed by a team internal to the organization being assessed (ISO/IEC 2004).

## 3. Methodology

In order to provide an overview on the current state of the art on software process self-assessment methods, we perform a mapping study. The mapping study aims at providing a broad review on existing reports on software process self-assessment methods, classifying them and describing their methodology and results. The research questions of this study focus on which software process self-assessment methods exist and what are their characteristics, especially with respect to their process reference models and measurement frameworks. We also analyze how these methods were developed and evaluated. As mapping studies use the same basic methodology as systematic literature reviews, this study follows an adaptation of the procedure proposed by Kitchenham (2007), Kitchenham et al. (2010) and Petersen et al. (2008) as presented in Figure 3.

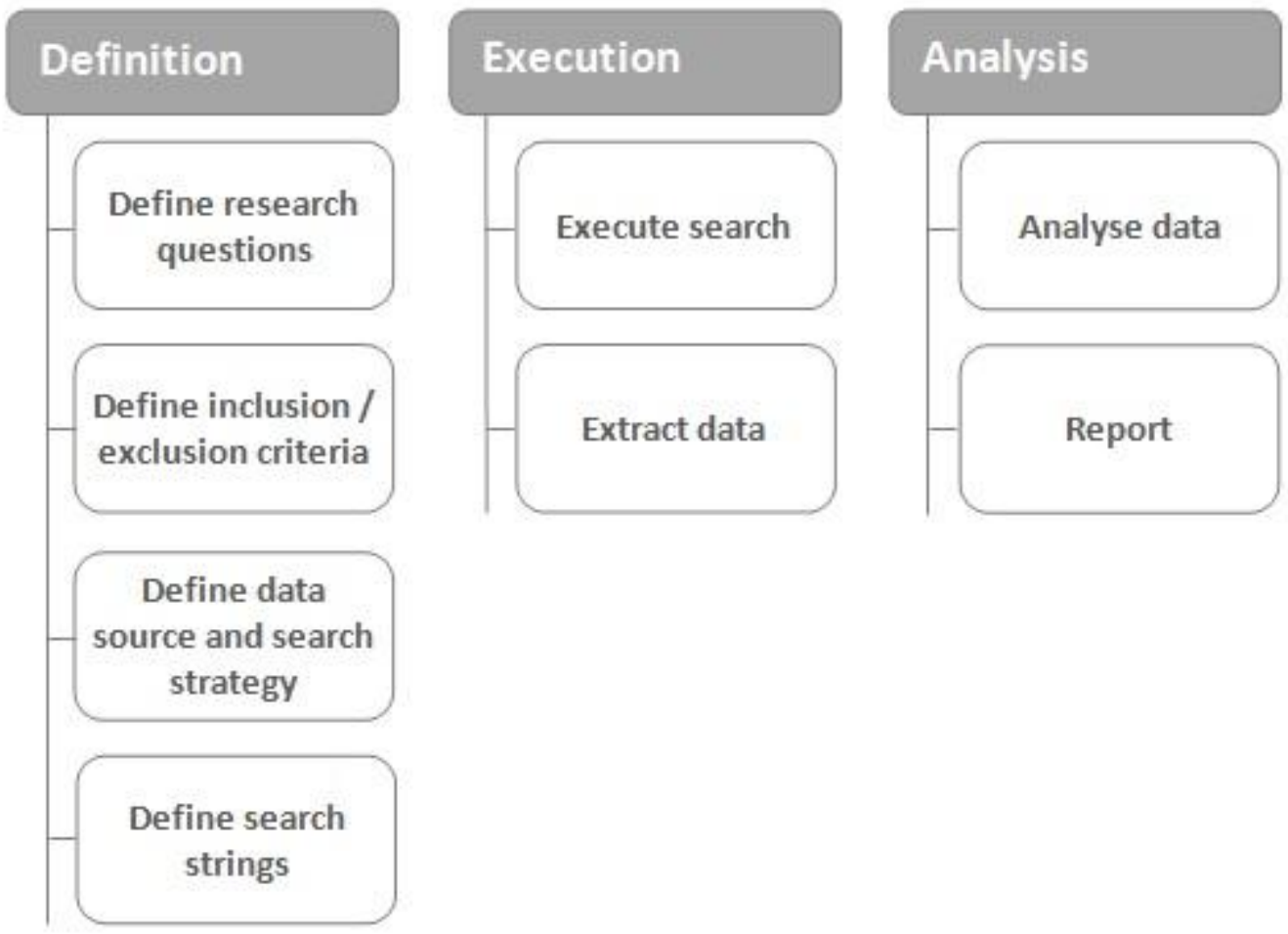

**Figure 3 Research methodology**

In the definition phase, the research questions and the review protocol were defined. The protocol contains the selection criteria to determine which studies will be included in the review, the data sources, search strategy and the definition of search strings.

The execution phase was carried out based on the review protocol conducting the search in the selected repositories. The initial search results were analyzed with respect to their relevancy applying the inclusion/exclusion and quality criteria.

Once identified the relevant studies, the data needed to answer the search questions was extracted. The extracted data are analyzed with respect to the defined research questions and the results are interpreted during the analysis phase.

## 4. Definition of the mapping study

The main research question driving this study is: What is the state of the art on software process self-assessment methods? Furthermore, we want to obtain an overview on the existing methods answering the following sub-questions:

RQ1. Which software process self-assessment methods exist?

RQ2. What are the characteristics of software process self-assessment methods in terms of assessment process, techniques and stakeholders?

RQ3. What are the characteristics of the process reference models?

RQ4. What are the characteristics of the measurement frameworks?

RQ5. Have the methods been developed and evaluated systematically?

**Data source and search strategy:** We examined all published English-language articles on software process self-assessment methods that are available on the Web via major digital libraries and databases in the computing field (IEEE Xplore, ACM Digital Library, Wiley, Springer and SCOPUS) with free access through the CAPES Portal[2]. To increase publication coverage, we also used Google Scholar, which indexes a large set of data across several different sources (Haddaway et al. 2015).

**Inclusion/exclusion criteria:** We only include studies, which present a software process self-assessment method, published in the last 24 years (starting at the initial release date of the CMM (Paulk 1993)) between January 1993 and June 2017. On the other hand, we excluded:

- Studies that present methods for the self-assessment of objects that are not software processes, such as products etc.
- Studies that developed a software process self-assessment method, but do not present the AM.
- Studies that present self-assessment methods for other contexts not related to software processes.
- Studies not written in English.

In case of duplicate reports of the same study, we considered the most recent complete report found.

**Quality criteria:** In addition to our inclusion/exclusion criteria, we also appraised the overall quality of the found studies. We considered only articles with substantial information on the process assessment method regarding our research questions.

**Definition of search string:** In order to calibrate the search string, we conducted informal searches in the aforementioned repositories. The string use in these searches contained combinations of expressions related to the research question, synonymous, related/broader concepts for each core concept synonyms, as well as abbreviations. The test strings contained combinations of the expressions self-assessment, “internal assessment”, software, “software process”, method, methodology, capability, maturity, guide and framework. As result of the calibration, we defined the generic search string “(self-assessment OR "internal assessment") AND process AND software AND ("maturity" OR "capability") NOT (students AND education)”.

| Repository | Search string |
|---|---|
| **Springer Link** | (self-assessment OR "internal assessment") AND process AND software AND ("maturity" OR "capability") NOT (students AND education)' within Engineering Computer Science English |
| **Wiley Online Library** | self-assessment OR "internal assessment" in All Fields AND process in All Fields AND software in All Fields AND "maturity" OR "capability" in All Fields NOT students AND education in All Fields |
| **ACM Digital Library** | (self-assessment OR "internal assessment") AND (process) AND (software) AND ("maturity" OR "capability") NOT (students AND education) |

[2]A web portal for access to scientific knowledge worldwide, managed by the Brazilian Ministry on Education for authorized institutions, including universities, government agencies and private companies (www.periodicos.capes.gov.br).

| | |
|---|---|
| **IEEE Xplore** | (((self-assessment OR "internal assessment") AND (process) AND (software) AND ("maturity" OR "capability") NOT (students AND education))) |
| **SCOPUS** | ALL ((self-assessment OR "internal assessment") AND (process) AND (software) AND ("maturity" OR "capability") AND NOT (students AND education)) AND PUBYEAR > 1992 |
| **Google Scholar** | "internal assessment" "software" "process" maturity capability -students -education |
| | "self-assessment" "software" "process" maturity capability -students -education |

**Table 1 Search string per repository**

Table 1 presents the search string used to perform the search in each repository.

# 5. Execution

## 2.2. Execution of search

The search has been realized in June 2017 by the first author and revised by the other authors (Table 2).

| Repositories | No. of initial search results | No. of articles analyzed | No. of articles selected after 1º stage | No. of articles selected after 2º stage |
|---|---|---|---|---|
| **Springer Link** | 189 | 189 | 6 | 0 |
| **Wiley Online Library** | 664 | 664 | 27 | 2 |
| **ACM Digital Library** | 337 | 337 | 28 | 0 |
| **IEEE Xplore** | 604 | 604 | 39 | 9 |
| **Google Scholar** | 222 (search string 1) 2040 (search string 2) | 900 | 142 | 27 |
| **SCOPUS** | 467 | 467 | 46 | 14 |
| **Total** | 4523 | 3383 | 287 | 33 (discounting 19 duplicates) |

**Table 2 Number of identified articles per repository per selection stage**

Table 2 presents the amount of articles found and reviewed in each repository.

In the first analysis stage, we quickly reviewed titles and abstracts to identify papers that matched the inclusion criteria, resulting in 287 articles potentially relevant. In the second stage, the articles were fully read with the objective to check their relevance with respect to our inclusion/exclusion criteria. In this step, 254 articles were excluded, most of them due to the fact that they address other forms of assessment than self-assessment, or deal with self-assessment of processes not related to software development. In this step, we also evaluated the articles with respect to the quality criteria. Some of the studies found (6) present case studies in which self-assessments were carried out in software companies or present the development of an assessment method, but do not provide enough details to analyze the AM, so they were also excluded. As result, 33 studies were considered relevant, as shown in Table 3.

| ID | Reference | Title |
|---|---|---|
| **1** | Garcia et al. 2010 | Adopting an RIA-Based Tool for Supporting Assessment, Implementation and Learning in Software Process Improvement under the NMX-I-059/02-NYCE-2005 Standard in Small Software Enterprises |
| **2** | Graden and Nipper 2000 | An Innovative Adaptation of the EIA/IS 731.2 Systems Engineering Capability Model Appraisal Method |
| **3** | Muladi and Surendro 2014 | The readiness self-assessment model for green IT implementation in organizations |
| **4** | Glanzner & Audy 2012[3] | 2DAM WAVE An Evaluation Method for the WAVE Capability Model |
| **5** | Widergren et al. 2010 | Smart grid interoperability maturity model |

[3] This study presents two assessment methods, however only the mini assessment version is considered in this research, as the extended assessment version requires the participation of an external assessor, and, therefore, cannot be performed by the organization in a totally autonomous way.

| | | |
|---|---|---|
| **6** | Burnstein et al. 1998 | A Model to Assess Testing Process Maturity |
| **7** | Grceva 2012 | Software Process Self-Assessment Methodology |
| **8** | Amaral & Faria 2010 | A Gap Analysis Methodology for the Team Software Process |
| **9** | Serrano et al. 2003 | An experience on using the team software process for implementing the Capability Maturity Model for software in a small organization |
| **10** | Bollinger & Miller 2001 | Internal capability assessments |
| **11** | Shrestha et al. 2015 | Evaluation of Software Mediated Process Assessments for IT Service Management<br>Building a Software Tool for Transparent and Efficient Process Assessments in IT Service Management |
| **12** | Blanchette & Keeler 2005 | Self Assessment and the CMMI-AM – A Guide for Government Program Managers |
| **13** | Wiegers & Sturzenberger 2000 | A Modular Software Process MiniAssessment Method |
| **14** | Yucalar & Erdogan 2009 | A Questionnaire Based Method for CMMI Level 2 Maturity Assessment |
| **15** | Kasurinen et al. 2011 | A Self-assessment Framework for Finding Improvement Objectives with ISO/IEC 29119 Test Standard |
| **16** | Karvonen et al. 2012 | Adapting the Lean Enterprise Self-Assessment Tool for the Software Development Domain |
| **17** | Varkoi 2010 | Process Assessment In Very Small Entities - An ISO/IEC 29110 Based Method |
| **18** | Shrestha et al. 2014 | Software-mediated process assessment for IT service capability management |
| **19** | Böcking et al. 2005 | A Lightweight Supplier Evaluation based on CMMI |
| **20** | Patel & Ramachandran 2009 | Agile maturity model (AMM): A Software Process Improvement framework for agile software development practices |
| **21** | Pino et al. 2010 | Assessment methodology for software process improvement in small organizations |
| **22** | Timalsina & Thapa 2016 | Assessment of software process improvement |
| **23** | Göbel et al. 2013 | Towards an agile method for ITSM self-assessment |
| **24** | Homchuenchom et al 2011 | SPIALS: A light-weight Software Process Improvement Self-Assessment Tool |
| **25** | Orci & Laryd 2000 | Dynamic CMM for Small Organisations - Implementation Aspects |
| **26** | Daily & Dresner 2004 | Towards Software Excellence |
| **27** | Coallier et al. 1994 | Trillium - Model for Telecom Product Development & Support Process Capability |
| **28** | MacMahon et al. 2015 | Development and validation of the MedITNet assessment framework: improving risk management of medical IT networks |
| **29** | Kar et al. 2012 | Self-assessment Model and Review Technique |
| **30** | Raza et al. 2012 | An open source usability maturity model (OS-UMM) |
| **31** | Rapp et al. 2014 | Lightweight Requirements Engineering Assessments in Software Projects |
| **32** | Abushama 2016 | PAM SMEs process assessment method for small to medium enterprises |
| **33** | Kuvaja et al. 1999 | TAPISTRY—A Software Process Improvement Approach Tailored for Small Enterprises |

**Table 3 Self-assessment methods found in the mapping study**

Table 3 presents the references and title of all publications selected.

## 2.3. Data extraction

We systematically extracted data from the articles in order to answer the research questions. Secondary sources (e.g., academic works) were also used to complete the information of the primary articles. In accordance to the research questions, we extracted the data described in Table 4 Data extracted from the studiesTable 4.

| **Research question** | **Data extracted** |
|---|---|
| RQ1. Which software self- | • Author(s) and title of the AM |
| assessment methods exist? | |
| R2. What are the characteristics of the methods in terms of assessment process, techniques and stakeholders? | • Activities (planning, data collection, data validation, process attribute rating and reporting)<br>• Technique used in each activity (interview, questionnaire, focus group, etc.)<br>• Tool support (templates, checklists, tables, etc.)<br>• Necessity for participants to have specific knowledge of SPI<br>• Customization for a specific domain<br>• Effort to perform a self-assessment |
| RQ3.What are the characteristics of the process reference models? | • SPCMM the reference model is based on<br>• Scope (process areas) |
| RQ4. What are the characteristics of the measurement frameworks? | • Measurement scale<br>• Calculation of results<br>• Questionnaire/checklist:<br>  ○ Amount of items<br>  ○ Item format and response scale<br>  ○ Examples/explanations<br>  ○ Respondents |
| RQ5. Have the methods been developed and evaluated systematically? | • Research design:<br>  ○ Development methodology<br>  ○ Evaluation methodology<br>  ○ Amount of data points<br>  ○ Evaluated characteristics<br>  ○ Evaluation context(s) |

**Table 4 Data extracted from the studies**

Table 4 presents the data extracted from the studies in order to answer the research questions. The extracted data is presented as part of the analysis of each of the research questions in the next section.

## 6. Data analysis

This section presents an analysis of the data extracted from the studies in accordance to the defined research questions.

### RQ1. Which software process self-assessment methods exist?

Thirty-three process self-assessment methods were found as listed in Table 3. The first study found proposing a software process self-assessment method was published in 1994. Figure 4 illustrates that the amount of studies publishing new software process self-assessment methods is increasing, which indicates that this topic is of continuous interest and still being researched.

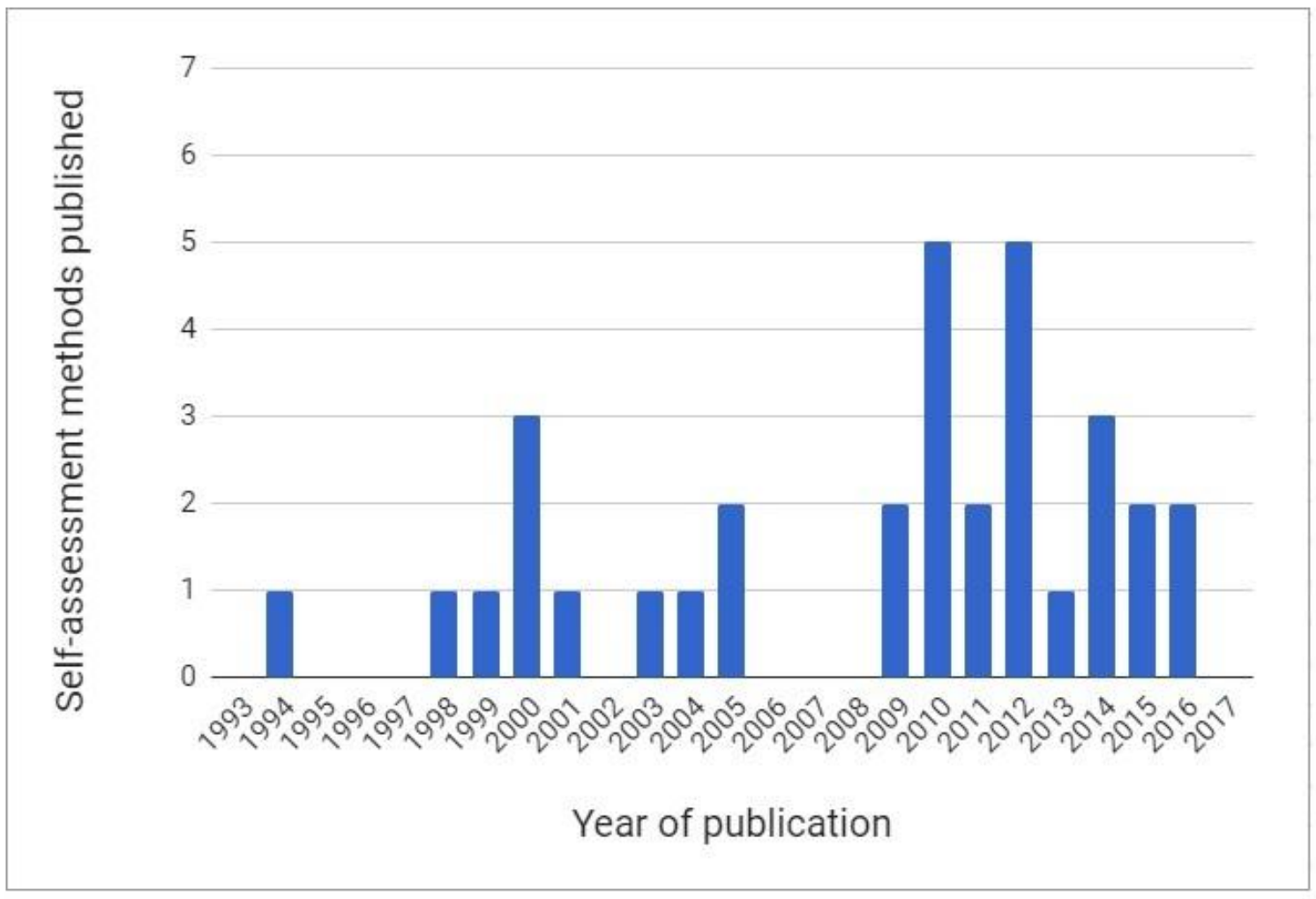

**Figure 4 Number of self-assessment methods per year of publication**

**RQ2. What are their characteristics in terms of assessment process, techniques and stakeholders?**

Analyzing this question, we observed that basically none of the published methods explicitly defines the process/steps to be performed in the self-assessment. An exception is the data collection that in some way is described by almost all methods. Some articles implicitly describe the assessment activities, for example, presenting tools to carry out data collection or explaining how data validation should be performed.

Regarding the planning activity, 3 AMs suggest holding a meeting with stakeholders to plan the following activities (Graden et al 2000; Wiegers et al 2000; Varkoi 2015). Some AMs develop documents in which the assessment roles, as well as, which processes will be assessed are defined (Burnstein et al. 1998; Widergren et al. 2010; Glanzner & Audy 2012; Muladi et al. 2014). The most used technique for data collection is via questionnaires, followed by interviews. To perform this activity, the self-assessment methods often provide software tools (Muladi et al. 2014; Shrestha et al. 2014; Pino et al. 2010; Daily & Dresner 2004; Malanga et al. 2015; Glanzner & Audy 2012; Burnstein et al. 1998; Timalsina and Thapa 2016; Kuvaja et al. 1999; Rapp et al. 2014) or electronic spreadsheets (Graden et al. 2000; Böcking et al. 2005).The approach of asking questions directly in an digital environment is considered a faster and more efficient data collection method compared with other methods such as interviews (Deutskens et al. 2006). Also, the fact that assessment activities can be automated, such as the generation of results, offers an efficiency gain, which can be translated into significant cost savings (Shrestha et al., 2014).

More specifically we encountered the following information on the assessment process:

**Planning:** Fourteen articles discuss the planning activity, but most of them do not mention the techniques used to perform it. Three articles propose to hold meetings with the assessors and those responsible for the processes and four mention the development of documents with pertinent information about the assessment.

**Data collecting:** Except for one method (Coallier et al. 1994), all methods present a data collection activity. Most of them use questionnaires (30 methods), 4 of which are used as script for interviews and 4 are used for performing workshops. Ten methods use interviews and five held workshops or focus group sessions. Only five articles mention performing data collection by gathering documents and other artifacts. Seven methods combine multiple techniques for data collection. Some, for example, combining the use of questionnaires and interviews (Graden et al., 2000; Muladi et al., 2014; Pino et al., 2010; Timalsina and Thapa, 2016; Burnstein et al. 1998), questionnaires with participants' discussion (Burnstein et al. 1998) or interviews with documents analysis (Rapp et al., 2014). Such a triangulation may be important in order to draw valid conclusions.

**Data validation:** Only 8 articles mentioned the validation of the collected data. Some assessment methods (Glanzner & Audy, 2012; Shrestha et al., 2014; Graden et al., 2000), use a software system to group the collected information and automatically indicate the validity of the information. Yet, in some cases no further information has been encountered on this issue. Glanzner & Audy (2012) and Shrestha (2014) do not present how the tool identifies inconsistency in the data. The tool used by Graden et al. (2000) checks for conflicting responses. Amaral et al., (2010) propose the accomplishment of interviews to validate the collected information (artifacts) and an intermediate presentation of the results to management. On the other hand, Pino et al. (2010) suggests that the assessor; in parallel to the interviews and questionnaires, collects information in order to validate participants' responses. In a similar way, Rapp et al. (2014) suggest to cross-check the initial interview results with collected documents.

**Process attributes rating:** Half of the methods (16) perform the process attribute rating as suggested by the standards on which they are based, such as ISO/IEC 15504, CMM/CMMI, among others. The other methods define their own process rating, which can be performed manually or automated (Glanzner & Audy, 2012; Burnstein et al. 1998; Grceva, 2012; Shrestha et al., 2014; Timalsina and Thapa, 2016; Daily & Dresner, 2004).

**Reporting:** Sixteen articles mention an activity regarding the report of the assessment results, for which 3 provide a template for reporting. One method (Glanzner & Audy, 2012) provides a software tool to record the lessons learned during the assessment. Other methods, besides generating a report; mention the presentation of the results to the managers (Glanzner & Audy, 2012; MacMahon et al. 2015, Rapp et al. 2014) and the realization of feedback sessions (Varkoi 2015). However, in general, the articles do not describe in detail the content and format of the reports.

Appendix A provides an overview of the data extracted on the activities and techniques that are defined/mentioned by each assessment method.

**Specific methods for certain contexts of use:** Twenty-four assessment methods aim at supporting the assessment of software processes in specific contexts, 10 of them in domains such as open source, usability, IT service, system engineering, Green IT, Smart grid, etc. Three methods are specific for agile development and 14 are specific for small businesses (two of which are customized to both contexts).

**SPI Knowledge:** Several articles do not mention the SPI expertise required for assessors to carry out the assessment. Eight articles mention that the assessment method can be applied even by a team without specific knowledge on process assessment. On the other hand, eight articles state the need for the assessor to have knowledge on SPI. Other studies report the possibility of conducting trainings, if necessary.

Appendix B presents the data extracted on the context in which each method is applied and its demands with respect to the assessors' knowledge on SPI.

**Effort to perform assessment:** Half of the assessment methods analyzed (15 methods) address the efficiency of the self-assessment method. However, only four of them present the average time to perform an assessment (Table 5). Except for Abushama (2016), the effort to perform the whole assessment is around 300 hours. Six articles present the average time for data collecting only. As shown in Table 6, data collection is expected to be a quick activity, performed in one or two work days. The other articles only comment that the developed method should require little effort to be applied (Burnstein et al. 1998; Blanchette 2005; Karvonen et al. 2012; Göbel et al. 2013).

| Article | Effort to perform self-assessment |
|---|---|
| Garcia et al. 2010 | Average of 385 hours |
| Graden and Nipper 2000 | About 4-5 weeks (~160-200 hours) |
| Kasurinen et al. 2011 | Lasts 390 hours |
| Abushama 2016 | About 24-34 hours |

**Table 5 Effort to perform assessment**

| Article | Effort for data collection |
|---|---|
| Varkoi 2010 | Average of 1 hour per questionnaire. |
| Shrestha et al. 2014 | Average of 4 hours per participant in software project activities.<br>Assessors spent a total of 48 hours. |
| Pino et al. 2010 | Average of 16 hours for small organizations (11 hours for the assessor advisor and 5 hours for the organization). |
| MacMahon et al. 2015 | Focus group lasting approximately 2 hours. |
| Abushama 2016 | Average of 2 hours for one interview (recommends interview a total of 1- 5 persons). An average of 12 hours for documents analysis. |
| Kuvaja et al. 1999 | 2 days (~16 hours). |

**Table 6 Effort for data collection**

### RQ3. What are the characteristics of the process reference models?

Most of the assessment methods are based on an already consolidated process reference model (28 methods), while some are based on models being developed as part of the specific assessment method (5 methods). Seven methods are based on ISO standards, 12 methods are based on CMM/CMMI, and 6 are based on CMM/CMMI-based models (PSP, TSP, TMM, TIM, MoProsoft) as summarized in Appendix C. Most of the self-assessment methods do not focus on specific software processes areas, except for Kar et al. (2012), Böcking et al. (2005), and Kuvaja (1999) including only processes considered important for SMEs.

### RQ4. What are the characteristics of the measurement frameworks?

Analyzing this question, we observed that most AMs perform the process attribute rating and use measurement scales based on well-established standards such as ISO/IEC 15504 and CMMI. Almost all methods use some type of questionnaire to assist in data collection. In general, the questionnaires are not extensive (having less than 50 items) using closed-ended questions. One third of them offers some kind of help, such as examples and explanations of the items.

Analyzing this question in more detail:

**Process attribute rating:** Most assessment methods do not define a specific activity for process attribute rating, however almost all of them present the way this should be done. Half of them use CMMI and ISO/IEC 15504 measurement scales and also calculate the assessment result the same way as suggested by these models. Four methods do not mention how their results are calculated and others propose different ways of calculating the maturity/capacity level, as presented in Appendix D.

**Measurement scale:** Half of the assessment methods use consolidated AMs scales such as CMM/CMMI (10 models) or ISO/IEC 15504 (5 models). Two methods use the scale of the TMM model and two the EIA 731 scale. Seven methods use different scales proposed specifically for the method. Six methods do not mention the use of measurement scales. Instead of presenting a level on a maturity or capability scale as a result of the assessment, some articles propose different way for providing feedback. For example, Karl et al. (2000) presents a "profiled set of guidance and, therefore, an overall view of the capability of their software/IT practices". Daily & Dresner (2004), in turn, presents opportunities for process improvement. Blanchette (2005) provides as assessment result a graph representing the processes and their respectively scores, in which the bars depict the range of scores for each process area. The self-assessment method proposed by Shrestha et al. (2014) aims to provide information that can drive improvement of IT service processes, rather than providing a capability level. Wiegers et al. (2000) is also more concerned with identifying appropriate improvement opportunities rather than maturity level ratings.

In total, 30 methods use questionnaires for data collection, as presented in Appendix D. Some questionnaires, on the other hand, are used to guide interviews and workshops (Amaral et al 2010; Yucalar & Erdogan 2009; Göbel et al. 2013; Orci & Laryd 2000; Rapp et al. 2014; Kuvaja et al. 1999).

**Amount of questionnaire items:** Most articles do not present the questionnaire and only 16 inform the amount of questionnaire items. Among the articles that provide this information, most use instruments with up to 50 items (4 methods) or only one item for each attribute/process indicator (4 methods). Only 7 instruments have more than 50 items. In one instrument, the number of items depends on the input provided by the organization and another proposes a questionnaire for each key process area.

**Item format/response scale:** Among the methods that use questionnaires, 4 do not mention the format of the items. Among the articles that provide this information only 2 use open-ended questions (of which 1 also used a questionnaire with closed-ended questions), 22 methods use closed-ended items, of which 6 are affirmations and not questions. In general, instruments with closed-ended items use response scales with an average of 4 categories (ranging from 2 to 7 categories). Few questionnaires use a Likert scale (Likert 1932), or a dichotomous scale (satisfying or not the respective item).

**Example/explanation:** Among the methods that use questionnaires and checklists, 11 provide information to aid in the interpretation of the questions/items as part of the questionnaire. Some methods provide explanations for each process area or just examples for the items, if necessary.

**RQ5. Have the assessment methods been developed and evaluated systematically?**

With regard to this research question, we analyze the methodology used to develop the AMs, as well as the methodology used for their evaluation/validation.

**Development methodology:** Among the encountered articles, only 6 mention how the assessment methods were developed. Yet, 2 do not use a formal development methodology (Karvonen et al. 2012; Orci & Laryd 2000). Karvonen et al. (2012) states that *LESAT for Software* was developed from an adaptation of the Lean Enterprise Self-Assessment Tool (LESAT). Key concepts of the model were initially identified and considered also valid in the context of software development. These concepts were adapted in order to use terminology more appropriate to the new context of use of the model. In addition, comments and examples relevant to the software development domain were included. Orci & Laryd (2000) used an approach that starts with proposing a new model, followed by its application in case studies, measuring, analyzing, and validating it. The guidelines to implement the model were developed by common sense and based on the experience of the author.

On the other hand, Göbel et al. (2013), Shrestha et al. (2014) and Shrestha et al. (2015) used the formal Design Science Research (DSR) method (or an adaptation). DSR is a method for developing artifacts that consist of 7 steps: problem identification, objectives of solution, design & development, demonstration, evaluation and communication. In addition to DSR other development methodologies were used. Shrestha et al. (2014) used the Goal-Question-Metric (GQM) approach to ensure that the measurement follows a transparent workflow of assessment activities, since this approach defines a measurement model for software metrics on three levels: goal (conceptual level), question (operational level) and metric (quantitative level). On the other hand, Kasurinen et al. (2011) present only the methodology used to develop the assessment model, but not the assessment method, and, consequently, are not included in Table 7, which lists the methodologies used for the development of the self-assessment methods.

| Article | Method development methodology |
|---|---|
| Karvonen et al. 2012 | "Adaptation of LESAT" |
| Orci & Laryd 2000 | “Proposing a new model, followed by its application in case studies, measuring, analyzing, and validating it.” |
| Göbel et al. 2013 | DSR |
| Shrestha et al. 2014 | DSR and GQM |
| Shrestha et al. 2015 | DSR |

**Table 7 Articles reporting the development methodology used**

**Evaluation methodology:** Although most of the articles do not present the development methodology, the majority (24) presents some form of evaluation of the developed assessment method (Appendix E). Among these, 7 articles do not present evaluation results related to the self-assessment method, but only regarding to the process reference model or the result of the assessment performed in some companies.

Seventeen articles report the evaluation of the assessment method. The most evaluated factors are efficiency, effectiveness and comfort (also represented by the term "usability" in some studies). Among these, 14 articles carried out case studies applying the methods in companies or with groups of professionals, who belong to the target audience of the assessment. The sample size varies between 1 and 24 companies; yet, nine studies were conducted with three companies or less.

In addition to the conduction of case studies, Burnstein et al. (1998) also conducted a review of the questionnaire by a panel of software engineering experts. Expert panels have also used by other studies as the only way to evaluate the method. MacMahon et al. (2015) carried out a review with 5 experts. Karvonen et al. (2012), on the other hand, compare the elements of the method with an analogue one proposed by Ericsson. Göbel et al. (2013) states that the method was tested iteratively, but do not inform how the tests were performed.

## 7. Discussion

A significant number of 33 software process self-assessment methods was encountered in the literature. The majority of them (24 methods) aims to assess the software processes in general (without focusing on a specified domain) and uses a version of the CMM/CMMI model or ISO/IEC 15504 as process reference model. Other self-assessment methods focus on a variety of specific domains such as IT service, Green IT, software testing, with a considerable number of 24 methods for specific contexts of use, including 15 self-assessment methods customized to SMEs and agile enterprises.

Almost all AMs use measurement scales (27 methods), most of them adopting the CMM/CMMI or ISO/IEC 15504 scale as is, calculating the maturity/capability level of the process in the same way as suggested by these models. Most AMs provide numerical results, such as a score, or assign a level on a scale. However, considering the primary objective of self-assessments on process improvement, some methods focus exclusively on providing improvement feedback on the strengths and weaknesses of the assessed process(es). One assessment model proposes to at least present the score for each assessed process, in order to identify "weak" processes, and thus assist in identifying the processes that need to be improved.

With respect to the assessment process, most of the articles focus on discussing data collection and do not approach the other activities in detail. However, the lack of guidance on how to plan the assessment, validate the collected data, calculate and generate the results may result in an inaccurate assessment result and/or difficult the application of these methods in practice.

The questionnaires are to be answered by different participants' roles (such as tester, developer, manager, area leader), and in some AMs are used to guide interviews and workshops. Considering the concern to carry out assessments as efficiently as possible, these data collection instruments are kept succinct. Several AMs use only 1 item for each practice, for example, with a total of up to 50 items; only 7 instruments are composed of more than 50 items. Some AMs are concerned in collecting data using more than one technique, for example, combining the use of questionnaires and interviews, questionnaires with participants' discussion or interviews with documents analysis. Such a triangulation may be important in order to increase the validity of the conclusions.

A small number of articles (8) mention the validation of the collected data. Among them, 3 perform validation by comparing the data collected through different methods. These situations might make sense in a context in which the organization is not seeking a certification in order to understand the capability of its processes. On the other hand, it is essential that the assessment results are reliable and effective in order to correctly guide software process improvement actions. Regarding the process attribute rating, half of the AMs perform the process attribute rating following standards, such as ISO/IEC 15504 or CMM/CMMI. Half of the articles mention the reporting of the assessment results, yet, most of them do not report how it should be performed. Three AMs provide a template to guide the preparation of a report, 3 suggest the presentation of results to stakeholders and one suggests the realization of a feedback session. In general, the process self-assessment methods lack explicit definitions of the process that should be used to conduct the activities of the process assessment.

A self-assessment method should not require specific knowledge of process assessment standards or process reference models as being conducted by internal staff, which might not have specialized knowledge nor experience in software process assessment. This feature is very important as it allows the applicability of these AMs in organizations with few resources, either without staff specialized in SPI or not able to invest in SPI training for their staff. However, only few AMs do not require the necessity of the assessment participants to have specific subject knowledge. In general, we observed that, although, the lack of SPI knowledge and experience of the internal assessors poses a significant threat to the validity of self-assessment, this issue is not addressed by the reported methods. To meet this requirement, some AMs just provide examples or tips on how to use the data collection instrument and how to analyze the collected data.

Regarding the format of the questionnaire items, we observed that most use closed-ended questions or affirmations. Using open-ended questions, on the other hand, might increase the assessment effort, as the provision of qualitative data involves a demanding process (Saunders et al. 2009). They also increase the complexity for the respondents, who are often non-experts on the subject being assessed and who may not know exactly how to answer the questions. For these reasons open-ended questions may not be an ideal option. For measuring dichotomous variables, closed-ended questions are also preferred, as possible answers can be easily precoded (Kazi 2012). Typically, closed-ended question instruments have response scales with an average of 4 ordinal points (ranging from 2 to 7 points). Despite the impact the format of the response scale may have on the complexity on answering of questionnaire and it results, none of the articles justified the scale used, nor discussed on whether the respondents would be able to distinguish between the categories used (for example, between partially or largely achieved).

The great amount of effort required to carry out a process assessment is often mentioned as a disincentive to companies that wish to have insight into their processes. In this respect, one of the main advantages in carrying out a self-assessment is the minimization of time and effort. Nevertheless, few articles mention this benefit. In general, the articles that evaluated the performance of the AMs in relation to their effort report that assessment lasts between 160 and 390 hours. Some articles, on the other hand, presented only the effort to perform data collection, and others considered only the time for responding the data collection instrument, or include the time for data analysis. As a result, the reported efforts are not comparable. Yet, in general, the total duration for data collection does take more than 3 working days (22 hours).

As AMs should be valid, reliable, and cost efficient, they need to be developed systematically (Gresse von Wangenheim et al. 2010; Simonsson 2007), and in a similar way, in order to acquire data properly, data collection tools need to be designed in such a way that they can measure aspects of interest (Kazi 2012). However, we observed that most of the AMs we encountered do not report information on their development. Only 3 articles present the use of systematic methods for the development of the AM, using Design Science Research and the Goal-Question-Metric approach. However, even these articles do not present a systematic method for the development of the measurement framework or the development of the data collection instruments. However, systematic development is important in order to assure the validity of the obtained results, especially, when using questionnaires that need to be designed carefully in order to comprehensively cover the object to be measured, while at the same time minimizing the number of items in order to minimize the data collection effort. Considering also the possible inexperience of the respondents with respect to SPI, it becomes essential to carefully revise the wording of the items (Kazi 2012; Saunders et al. 2009). Although 17 articles have reported some form of evaluation of the AM, most of them conduct studies using only small samples and/or without using a systematic method. Yet, again, as most AMs use some kind of questionnaire, it is important to assure the reliability and construct validity of these instruments (Kazi 2012; Saunders et al. 2009). However, we observed that only four studies evaluated the questionnaire/checklist used. Thus, the lack of scientific rigor of the validation of the majority of the proposed methods may leave their validity questionable. In this context it would be important to conduct larger empirical studies in order to confirm the validity and reliability of the proposed methods, e.g. in the way this has been done as part of the SPICE Project (Jung et al. 2001).

**Threats to Validity**

As with all research, there exist several threats to the validity of the results presented (Zhou et al. 2016). We, therefore, identified potential threats and applied mitigation strategies in order to minimize their impact on our research. A main risk in a mapping study is the omission of relevant studies. In order to mitigate this risk, the search string was carefully constructed and verified by the authors. Different strings (containing the core concepts and their synonyms) were tested in order to identify the one that returned the most relevant results (including seminal articles). The risk of excluding a relevant study is further mitigated by the use of multiple repositories that cover many repositories. In addition, we used Google Scholar as a complementary repository to find grey literature. Although, the use of Google Scholar as a single source for performing mapping studies is not recommended, its use as a complementary repository is reported in several articles (Haddaway et al. 2015). In accordance to Haddaway et al. (2015), who suggest that the first 200-300 results on Google Scholar should be revised, we analyzed all results using the first string (22 search results) and the first 900 search results for the second search string.

Threats to study selection and data extraction have been mitigated through a detailed definition of the inclusion/exclusion criteria. We defined and documented a rigid protocol for the study selection and all authors conducted the selection together discussing the selection until consensus was achieved. Recognizing the lack of consistent use of terminology, the information of the encountered articles has been carefully extracted and revised interpreting the presented models in relation to the theory presented in the background section. In this respect, this article presents an interpretative analysis with findings based on the author's subjective interpretations. In order to reduce risks of misunderstanding again all authors conducted the data extraction together until consensus was achieved.

## 8. Conclusion

We encountered a considerable amount of 33 software process self-assessment methods. Most methods are based on traditional and consolidated process reference models and measurement frameworks, such as CMMI and ISO/IEC 15504. On the other hand, in relation to these traditional models, the assessment process, in general, is simplified. Half of the AMs are customized for SMEs, which require less costly, more efficient and less bureaucratic assessment methods. Most of the AMs perform data collection through the application of (closed-ended) questionnaires. Benefits of the data collection technique are a reduction of time and effort and the ease to analyze the data as well as the possibility of automating the generation of the results. However, although, several studies mention concerns about the duration and effort of the assessment, very few analyzed these factors of the proposed AM. Another shortcoming observed is that with respect to the assessment process, most methods focus exclusively on data collection not detailing other activities. This may complicate the application of the proposed AMs in practice. We also observed that besides few exceptions, the methods for developing these AMs have not been reported. And, very few articles report the validation of the methods and/or the data collection instruments. Another issue observed is that most proposed AMs do not discuss the mitigation of significant threats to the validity of the results obtained through self-assessment due to the potential inexperience of the assessors and the lack of data triangulation. Therefore, it seems questionable as to whether the majority of the proposed AMs can be used to perform valid self-assessments. In this respect, this article provides an overview on existing software process self-assessment methods and their characteristics. It also points out important issues with respect to the development of new AMs and/or the improvement of existing ones as well as the need for the conduct of more rigorous larger-scale validations as a basis for future research in this area.

**Acknowledgements**

We would like to thank Prof. Stan and Martha Hanson for the review of this article.

This work was supported by the CNPq (*Conselho Nacional de Desenvolvimento Científico e Tecnológico – www.cnpq.br*), an entity of the Brazilian government focused on scientific and technological development.

## Appendix A – Overview on the assessment methods activities

Table 8 provides an overview on the activities that are defined/mentioned in the studies and indicate the techniques proposed to carry out them. The symbol "*" indicates that the study contains the stage but do not mention its practices. Information that is not provided in the articles is indicated by "-".

| Article | | Assessment activities | | | | |
|---|---|---|---|---|---|---|
| | | Planning | Data collection | Data validation | Process attribute rating | Reporting |
| Garcia et al., 2010 | Activity | - | * | - | * | - |
| | Technique | - | Questionnaire | - | ISO/IEC 15504 | - |
| Graden and Nipper, 2000 | Activity | * | * | * | * | * |
| | Technique | Discussions with the consultant and with the target group leadership | Questionnaire (in a spreadsheet) and interview | The questionnaire was imported into the database for validation. Charts served to provide insight about internal differences within the target group. | EIA/IS 731.2 | - |
| Muladi and Surendro, 2014 | Activity | * | * | - | * | - |
| | Technique | Determination of readiness factors, interview questions and questionnaire. | Interview and questionnaire (software) | - | ISO/IEC 15504 | - |
| Glanzner & Audy, 2012 | Activity | * | * | * | * | * |
| | Technique | Requirements Analysis, Selection and Preparation of the Team and Development of the Evaluation Plan (Document) | Questionnaire (software) | The tool calculates if the data collected are valid, or if there was any relevant discrepancy between the results through a heuristic, | Mini: The responses indicate the level of each attribute, and at the end of the questionnaire, the unit capability level. Extensive: Preparing Participants, Evidence and Affirmations Collection, Evidence and Affirmations Documentation, Evidence and Affirmations Verification, Validate the First Discoveries. | All relevant artifacts are included in the WAVE's database of historical Data. The leader of the assessment presents to all stakeholders. |
| Widergren et al., 2010 | Activity | * | * | - | * | - |
| | Technique | Define target, domain, goals and identify practices. (Assessment and planning process diagrams). | Gather evidence and others and use maturity model tools | - | CMMI | - |
| Burnstein et al., 1998 | Activity | * | * | - | * | * |
| | Technique | A statement of assessment purpose, scope, and constraints is prepared to guide the development of the assessment plan. | Interviews, presentations, questionnaires (soft. tool), and relevant documents. | - | The ranking algorithm requires a rating of the maturity subgoals, then the maturity goals, and finally the maturity level. | The profile can be presented as a graphical display or in the form of a matrix that indicates maturity goals that are satisfied or not, the TMM level, a summary of test process strengths and weaknesses, and recommendations for improvements. |
| Grceva, 2012 | Activity | - | * | - | * | - |
| | | - | Document-based | - | Documents are inserted in a soft. tool that statistically generates the analysis of the results. | - |
| Amaral & Faria, 2010 | Activity | * | * | * | * | * |
| | Technique | - | Interviews (based | Perform interviews to | ISO/IEC 15504 | Report template. |
| | | | on questionnaires). | validate the collected information (artifacts) | | |
| Serrano et al., 2003 | **Activity** | - | * | - | * | - |
| | **Technique** | | Questionnaire | | CMM | |
| Bollinger & Miller, 2001 | **Activity** | - | * | - | * | - |
| | **Technique** | | Questionnaire | | EIA/IS-731 | |
| Shrestha et al., 2015 | **Activity** | - | * | - | * | - |
| | **Technique** | - | Questionnaire (Developed tool) | | ISO/IEC 15504 | |
| Blanchette & Keeler, 2005 | **Activity** | - | * | - | Does not apply | - |
| | **Technique** | - | Questionnaire | - | - | - |
| Wiegers & Sturzenberger, 2000 | **Activity** | * | * | - | * | * |
| | **Technique** | The assessors meet with the project's software leader to plan the activities. | Questionnaire and Participant discussion | - | The assessors analyze the questionnaire responses using a spreadsheet tool. | 1. Assessors present findings to project team<br>2. Project team presents findings to their management. |
| Yucalar & Erdogan, 2009 | **Activity** | - | * | - | - | - |
| | **Technique** | - | Interview (based on questionnaire). | - | - | - |
| Kasurinen et al., 2011 | **Activity** | - | * | - | - | - |
| | **Technique** | | Rounds of Interviews (questionnaire based) | | | |
| Karvonen et al., 2012 | **Activity** | - | * | - | - | - |
| | **Technique** | - | Checklist | - | - | - |
| Varkoi, 2010 | **Activity** | * | * | * | * | * |
| | **Technique** | Face-to-face meetings. | Interview | - | ISO/IEC 15504 | Feedback sessions |
| Shrestha et al., 2014 | **Activity** | - | * | * | * | * |
| | **Technique** | - | Questionnaire (Developed tool) | Developed software tool calculates the coefficient of variation score. | Developed software tool. | The tool extracts a recommendation item from the knowledge base and the items are compiled into an assessment report. |
| Böcking et al., 2005 | **Activity** | - | * | - | * | * |
| | **Technique** | - | Questionnaire (Excel application) | - | CMMI | Delivers the supplier the results of the evaluation and plans steps for improvement |
| Patel & Ramachandran, 2009 | **Activity** | - | * | - | - | - |
| | **Technique** | - | Questionnaire | - | - | - |
| Pino et al., 2010 | | * | * | * | * | * |
| | **Technique** | - | Interview and survey (EvalTool). | The assessor gathers information separately from the person responsible for the process to be assessed so the documentation of the organization's processes is inspected. | ISO/IEC 15504 | Assessment report template. |
| Timalsina & Thapa, 2016 | **Activity** | - | * | - | * | - |
| | **Technique** | - | Questionnaires (soft. tool) and interviews. | - | Using a software tool to determine the final score for each capability level by calculating the mean value of all the responses for that level by all the respondents. | - |
| Göbel et al., 2013 | **Activity** | * | * | * | - | * |
| | **Technique** | - | Workshop (based on checklist) | - | When all participants understand the meaning of the statement the | Notes and graphs |
| | | | | | group discusses the different ways they work and agree on a "rating". Additional metrics (optional) are selected as complement to the self-assessment rating. | |
| Homchuenchom et al, 2011 | **Activity** | * | * | - | * | * |
| | **Technique** | - | Questionnaire | - | CMMI | - |
| Orci & Laryd, 2000 | **Activity** | * | * | - | * | * |
| | **Technique** | Select - Appoint - Train: involves both selection of the appropriate model, appointment of people to I-Roles, and training. | Workshop (based on checklist) | - | A defined and documented process must be approved by the working group and SEPG. If an approved status cannot be directly reached, a new workshop should be arranged. | Documented D-process |
| Daily & Dresner, 2004 | **Activity** | - | * | - | * | * |
| | **Technique** | - | Questionnaire (web based tool) | - | Responses are combined using a weighting scheme. This weighting takes into account the number of questions on the form and the importance of each one. | A standard format of Assessment Report is available. |
| Coallier et al., 1994 | **Activity** | - | - | - | * | - |
| | **Technique** | - | - | - | - | - |
| MacMahon et al., 2015 | **Activity** | - | * | - | * | * |
| | **Technique** | - | Focus group interviews (based on questionnaires) | - | ISO/IEC 15504 | A report document is generated and presented. |
| Mesquita and Barros, 2014 | **Activity** | - | * | - | * | - |
| | **Technique** | - | Questionnaire | - | CMM | - |
| Kar et al., 2012 | **Activity** | - | * | - | * | - |
| | **Technique** | - | Questionnaire | - | SMART SPICE | - |
| Raza, etal., 2012 | **Activity** | - | * | - | * | - |
| | **Technique** | - | Questionnaire | - | The maturity is determined by the extent to which the project managers and developers agree with each statement in the questionnaire. | - |
| Rapp et al., 2014 | **Activity** | * | * | * | * | * |
| | **Technique** | Kick-off meeting inclusive handing over the requirements engineering artifacts. | Initial Document Analysis and/or Interviews (based on questionnaire in a soft. tool) | Cross-checking initial interview results and documentation. | Do not apply | Final report with diagrams and presentation to the process representatives. |
| Abushama, 2016 | **Activity** | * | * | - | * | * |
| | **Technique** | - | Questionnaire | - | SCAMPI C | List of improvement areas. |
| Kuvaja et al., 1999 | **Activity** | - | * | - | * | - |
| | **Technique** | - | Workshop (BootCheck tool) | - | ISO/IEC 15504 | - |

**Table 8 Self-assessment activities and techniques**

## Appendix B – Context of use and required knowledge in SPI of the assessment methods

| Article | Context of use | Knowledge in SPI |
|---|---|---|
| Garcia et al., 2010 | - | Requires knowledge in SPI |
| Graden and Nipper, 2000 | SME enterprises | Does not require. |
| Muladi and Surendro, 2014 | Green IT | - |
| Glanzner & Audy, 2012 | Global software development | - |
| Widergren et al., 2010 | Smart grid | Requires knowledge in SPI |
| Burnstein et al., 1998 | Software testing | Requires knowledge in SPI |
| Grceva, 2012 | - | - |
| Amaral & Faria, 2010 | - | Requires knowledge in SPI |
| Serrano et al., 2003 | SME enterprises | - |
| Bollinger & Miller, 2001 | - | - |
| Shrestha et al., 2015 | IT service | - |
| Blanchette & Keeler, 2005 | Acquisition | Does not require. |
| Wiegers & Sturzenberger, 2000 | - | Does not require. |
| Yucalar & Erdogan, 2009 | SME enterprises | - |
| Kasurinen et al., 2011 | Software testing | - |
| Karvonen et al., 2012 | Lean | - |
| Varkoi, 2010 | SME enterprises | Does not require. |
| Shrestha et al., 2014 | IT service | - |
| Böcking et al., 2005 | SME enterprises | Does not require (only a basic understanding of CMMI) |
| Patel & Ramachandran, 2009 | SME enterprises and agile | - |
| Pino et al., 2010 | SME enterprises | Requires knowledge of the methodology and assessment of the application process, and analysis of the data collected. |
| Timalsina & Thapa, 2016 | - | - |
| Göbel et al., 2013 | Services | - |
| Homchuenchom et al, 2011 | SME enterprises and agile | - |
| Orci & Laryd, 2000 | SME enterprises | Requires at least a leader knowledge of both SPI and software development. |
| Daily & Dresner, 2004 | SME enterprises | - |
| Coallier et al., 1994 | - | Requires knowledge in SPI |
| MacMahon et al., 2015 | Risk management | - |
| Kar et al., 2012 | SME enterprises | Requires knowledge in SPI |
| Raza et al., 2012 | Open source usability | Does not require. |
| Rapp et al., 2014 | - | Does not require knowledge in SPI, but in RE. |
| Abushama, 2016 | SME enterprises | Does not require. |
| Kuvaja et al., 1999 | SME enterprises | Requires knowledge in SPI |

**Table 9 Contexts of use and SPI knowledge requirements**

## Appendix C - Source on which the SPCMMs are based

| Article | SPCMM based on | Scope |
|---|---|---|
| Garcia et al., 2010 | NMX-I-059/02-NYCE-2005 (Moprosoft) | All process areas |
| Graden and Nipper, 2000 | EIA/IS 731.2 Systems Engineering Capability Model Appraisal Method | All process areas |
| Muladi and Surendro, 2014 | Green IT implementation (developed by the author) | All process areas |
| Glanzner & Audy, 2012 | WAVE capability model | All areas: people, projects, unit and portfolio |
| Widergren et al., 2010 | Smart grid interoperability maturity model (developed by the author) | All areas Configuration & Evolution, Operation, Security & Safety |
| Burnstein et al., 1998 | TMM | 16 process areas |
| Grceva, 2012 | CMM | All process areas |
| Amaral & Faria, 2010 | Team Software Process (TSP) | All process areas |
| Serrano et al., 2003 | SW-CMM,TSP, PSP. | All process areas |
| Bollinger & Miller, 2001 | EIA/IS-731 | All process areas |
| Shrestha et al., 2015 | ISO 20000 and IT Infrastructure Library (ITIL) | All process areas |
| Blanchette & Keeler, 2005 | CMMI-AM | All process areas |
| Wiegers & Sturzenberger, 2000 | CMM | All process areas |
| Yucalar & Erdogan, 2009 | CMMI | All process areas |
| Kasurinen et al., 2011 | ISO/IEC 29119 and TIM | Similar to ISO/IEC 29119 processes organization is conceptually close to organizational management process (OTP), planning and tracking to test management process (TMP) and TMCP, test cases to test plan process (TPP), test ware to STP and DTP, and reviews to TCP. |
| Karvonen et al., 2012 | Lean Enterprise Model (LEM), the enterprise Transition-To-Lean (TTL) roadmap. | All process areas |
| Varkoi, 2010 | ISO/IEC 29110 | All process areas |
| Shrestha et al., 2014 | ISO/IEC TR 20000-4:2010 | All process areas |
| Böcking et al., 2005 | CMMI | Process areas from level 2 and 3 |
| Patel & Ramachandran, 2009 | AMM model (developed by the author) | All process areas |
| Pino et al., 2010 | PmCOMPETISOFT | All process areas |
| Timalsina & Thapa, 2016 | CMMI | All process areas |
| Göbel et al., 2013 | CMMI-SVC 1.3 e ARC | All process areas: Strategic Service Management (STSM), Service System Development (SSD), Service System Transition (SST), Service Delivery (SD), Incident Resolution and Prevention (IRP), Capacity and Availability Management (CAM), Service Continuity (SCON) |
| Homchuenchom et al, 2011 | CMMI + SCRUM | Process areas Project Planning (PP), Project Monitoring and Control (PMC) and Integrated Project Management (IPM) |
| Orci & Laryd, 2000 | CMM | All process areas |
| Daily & Dresner, 2004 | TSE Model (developed by the author) | All process areas: Customer - Supplier, Engineering, Support, Management, |
| | | Organisation, Legal |
| Coallier et al., 1994 | Trillium (developed by the author) | All process areas |
| MacMahon et al., 2015 | IEC 80001-1 | All process areas |
| Kar et al., 2012 | ISO/IEC 12207 | Basic dimension based on ISO/IEC 12207 |
| Raza et al., 2012 | OS-UMM (developed by the author) | All process areas |
| Rapp et al., 2014 | Requirements Engineering Reference Model (REM) and Requirements Capability Maturity Model (R-CMM) | All process areas |
| Abushama, 2016 | CMMI | All process areas |
| Kuvaja et al., 1999 | BOOTSTRAP 3.0 | 19 processes were considered to include the most important processes for any SMEs |

**Table 10 Characteristics of the reference models**

## Appendix D - Characteristics of the assessment methods' measurement framework

Table 11 presents the characteristics of the methods regarding their measurement framework and their data collection instruments, such as, process attribute rating, measurement scale, amount of items and response scale. Methods that do not present the questionnaire itself are indicated by "-". Methods that do not use questionnaires are indicated by "N/A".

| Article | Process attribute rating | Measurement scale | Amount of items | Items format | Response scale | Provides example/explanations | Respondent |
|---|---|---|---|---|---|---|---|
| Garcia et al., 2010 | Adopting ISO/IEC 15504 | ISO/IEC 15504 capability levels | - | Closed | 7 point ordinal scale: Always, Usually, Sometimes, Rarely if ever, and Never. Don't Know and Not Apply. (Comments) | - | Project managers |
| Graden and Nipper, 2000 | Adopting EIA/IS 731.2 | EIA/IS 731.2 capability levels | More than 600 items | Closed | 3 point nominal scale: Yes, no, don't know, not apply. | Explanation for each question session. | Program/project management |
| Muladi and Surendro, 2014 | Adopting ISO/IEC 15504 | Readiness levels 0-50 Not ready 51-85 Ready 85-100 Prepared | - | - | - | - | Organization's manager |
| Glanzner & Audy, 2012 | The levels of implementation are defined based on the number of evidence e=and weak points found. "Fully Implemented" and "Largely Implemented" indicates that the practice was considered implemented. | Capability levels 2, 3, 4 | One item for each model attribute (26 attributes). | - | - | - | Two groups of professionals, three with technical responsibilities and three with management responsibilities. |
| Widergren et al., 2010 | Adopting CMMI | CMMI maturity Levels | N/A | N/A | N/A | N/A | N/A |
| Burnstein et al., 1998 | The ranking algorithm requires a rating of the maturity subgoals, then the maturity goals, and finally the maturity level. | TMM scale: Level 1: Initial Level 2: Phase definition Level 3: Integration Level 4: Management and Measurement Level | - | - | - | Instructions for use. Recommendations for questionnaire improvement. A glossary of testing terms | - |
| | | 5:Optimization, defect, prevention and Quality control. | | | | | |
| Grceva, 2012 | Adopting CMM (soft. tool) | CMM Levels | N/A | N/A | N/A | N/A | N/A |
| Amaral & Faria, 2010 | - | - | 55 for managers<br>46 for testers<br>61 for developers | Open | - | - | Managers, developers and testers. |
| Serrano et al., 2003 | Adopting CMM | CMM Levels | 124 items (mean of 6-7 per KPA) | Closed | 4 point nominal scale:<br>Yes, No, Does Not Apply, and Don't Know. | Instructions for use and explanation for each KPA and main concepts | - |
| Bollinger & Miller, 2001 | Adopting EIA/IS-731 | EIA/IS-731 capability Levels | - | Closed | 3 point nominal scale: Yes, no, not apply. A "yes" answer required a brief comment that cited an example of the type of activity that was carried out that met the practice. A "NA" answer required a brief comment to justify why it was not applicable. | Examples for some items. | A mix of participants that gave sufficient coverage of the various engineering disciplines as well as a mix of practitioners and leaders. |
| Shrestha et al., 2015 | Adopting ISO/IEC 15504 | ISO/IEC 15504 capability levels | - | Closed | 5 point ordinal scale:<br>No, Partially, Largely, Fully and Not Applicable | - | Process stakeholders |
| Blanchette & Keeler, 2005 | - | - | 30 affirmations. Questions covering all the process areas described in the CMMI-AM. | Closed | 10 point ordinal scale.<br>Score each statement from 1 to 10. Statement could be positive or negative. | - | Program manager and deputy program manager, chief engineer, chief software engineer, contracts specialist, business manager, and leads of integrated product teams. |
| Wiegers | - | - | - | Closed | 7 point ordinal | No. | Organization |
| & Sturzenberger, 2000 | | | | | scale: Always, Usually, Sometimes, Rarely, Never, Don't Know, Not Applicable | The assessors facilitate the questionnaire administration session, using standard slides to describe the intent of each KPA before the participants answer the questions for that KPA. | representatives |
| Yucalar & Erdogan, 2009 | A score of 80 or better, most likely indicates having achieved the maturity level 2. | CMMI maturity level | 39 items. | Closed | 5 point ordinal scale: definitely yes, usually, planned but not applied, not sure, definitely no | - | A responsible and knowledgeable person. |
| Kasurinen et al., 2011 | Compare the observations made with a profile that indicates the maturity level. | TIM levels; Level 0, Initial, Level 1, Baseline, Level 2, Cost-effectiveness, Level 3, Risk-lowering, Level 4, Optimizing. | - | - | - | - | Software designer, test manager, manager. |
| Karvonen et al., 2012 | The capability level is decided upon by using lean indicators and capability-level descriptions for the specific practice. | Least capable (Level 1) to world class (Level 5). | 54 affirmations (one for each practice) | Closed | 2 point nominal scale: C (current) and D (desired). | Examples of indicators for each practice. | Leadership of the enterprise |
| Varkoi, 2010 | Adopting ISO/IEC 15504 | ISO/IEC 15504 capability levels | N/A | N/A | N/A | N/A | N/A |
| Shrestha et al., 2014 | - | - | One for each indicator. | Closed | 5 point ordinal scale: Not, Partially, Largely, Fully and Not Applicable | - | Process performers, process managers and other process stakeholders |
| Böcking et al., 2005 | Adopting CMMI | CMMI Levels | One sheet (questionnaire) for each of the seven process areas, one sheet for the assessment details, and | Statement | 3 point ordinal scale: Yes, Partially, No | Each questionnaire begins with a short description of the process area. | - |
| | | | one sheet for the presentation of results. | | | | |
| Patel & Ramachandran, 2009 | The answer of the questionnaires is used as input in a formula that calculates a percentage of achievement for each PAs. The degrees of achievement are: Fully Achieved, Largely Achieved, Partially Achieved and Not Achieved. KPA identifies the issues that must be addressed to achieve a maturity level. | Maturity levels: Initial Explored, Defined, Improved. Sustained. | 94 affirmations (mean of 7 items per PA) | Closed | 4 point nominal scale: Yes, Partially, No, Not Applicable (N/A) | - | Developers, coach, testers with collaboration of on-site customer. |
| Pino et al., 2010 | Adopting ISO/IEC 15504 | ISO/IEC 15504 capability up to level 2. | - | Statement | Ordinal scale. Assigned a numeric value of 0 (never), 0.5 (sometimes) or 1 (always). | Each Process attribute begins with a short description. | - |
| Timalsina & Thapa, 2016 | Adopting CMMI. The software tool determines the final score for each capability level by calculating the mean value of all the responses for that level by all the respondents. | CMMI capability levels | 36 items. | Open and closed | 4 point nominal scale: Never, Sometimes, Almost & Always. | - | Key resource person of the company. |
| Göbel et al., 2013 | Adopting CMMI-SVC | CMMI-SVC 1.3 Levels | - | - | - | - | - |
| Homchuenchom et al., 2011 | Practices are scores in strength, weak and not rated. The practices indicator determines the practice characteristic that in turn indicates the goal satisfaction that indicates the process area satisfaction. | - | Questionnaire generated based on the organization's input | Statement | 3 point nominal scale: (a) Use, (b) Do Not Use, and (c) Not Available to Use | - | Each organization's project role provides evidences, related with its role. |
| Orci & Laryd, 2000 | Adopting CMM | CMM up to level 2. | One item for each activity from each PA. | Statement | 2 point nominal scale: Checked, not checked | - | Senior Manager, Project Manager, SoftWare Manager, Software Engineering group, SQA group |
| Daily & Dresner, 2004 | - | - | Sets of questions for each practice that is going to be assessed. | Closed | 6 point ordinal scale: always, usually, sometimes, rarely or never (or is not applicable) or 3 options: yes or no (or is not applicable). | Explanation about the process area. Examples for some items. | - |
| Coallier et al., 1994 | To achieve a level, an organization must satisfy a minimum of 90% of the criteria in each of the 8 Capability Areas at that level. Levels 3, 4 and 5 require the achievement of all lower levels | Capability levels: 1. Unstructured, 2 Repeatable and Project Oriented, 3 Defined and Process Oriented 4 Managed and Integrated, 5 Fully Integrated. | - | - | - | - | - |
| MacMahon et al., 2015 | Adopting ISO/IEC15504 | ISO/IEC 15504 capability levels | - | - | - | - | Risk management stakeholders |
| Kar et al., 2012 | Each question is awarded with score range 0-5, so the maximum score is “125” point and the minimum score is “0”. | Capability levels: 0 to 50% - Poor, 51% to 65% - Fair, 66% to 80% - Average & manageable, 81% to 90% - Established above 90% - Well established | 125 items, 5 per process area | Closed | 6 point ordinal scale: 0-Not attained at all, 1- poorly attained, 2- fairly attained, 3- attained averagely 4- Largely attained, 5- completely attained | - | - |
| Raza et al., 2012 | Presents a formula to calculate the Usability Maturity Level based on the questionnaires responses. | Maturity levels: 1: Preliminary 2: Recognized 3: Defined 4: Streamlined 5: Institutionalized | 111 affirmations | Statement | 5 point ordinal scale: Fulfilled, Largely Fulfilled, Partially Fulfilled, Not Fulfilled and Not Applicable. | - | Project managers or developers. |
| Rapp et al., 2014 | - | - | - | Closed | 4 point ordinal scale: Definitely yes, Rather yes, | - | A company representative who have |
| | | | | | Rather no, no. | | been involved in the RE activities. |
| Abushama, 2016 | KPA satisfaction level (not achieved, partially achieved, largely achieved and fully achieved) is calculated through a formula. | CMMI levels | A questionnaire for each KPA | Closed | 4 point ordinal scale: Yes–Partially–No–Does Not Apply | - | Certified/experienced assessor by the SEI. |
| Kuvaja et al., 1999 | Adopting ISO/IEC 15504 | ISO/IEC 15504 capability levels | - | - | - | - | - |

**Table 11 Characteristics of the measurement frameworks**

## Appendix E – Assessment methods evaluation

Table 12 presents information on how the assessment methods have been evaluated, which characteristics were assessed, in which context and with how many data points.

| Research design | Articles | Evaluated characteristics | Evaluation context | Amount of data points |
|---|---|---|---|---|
| **Case study** | Shrestha et al., 2015 | Effectiveness, Efficiency, Usefulness, Trust, Comfort | Customer contact management, spatial information services for improved web mapping services, including mobile solutions, etc. | 1 company, 9 participants |
| | Graden and Nipper, 2000 | Effectiveness | Department of Energy field office | 14 people |
| | Grceva, 2012 | - | - | 1 company |
| | Wiegers & Sturzenberger, 2000 | Cost, Efficiency, user satisfaction | Small projects. Average project team size was 12. | 24 projects |
| | Kasurinen et al., 2011 | Accuracy and usability. | A large, internationally operating software company. A small-sized company, producing solutions for customer organizations. A large company producing software used for computer-assisted design. A medium-sized company, producing embedded software. | 3 companies |
| | Shrestha et al. (2014) | Transparency and efficiency (the degree of economy with which any assessment consumes resources, especially time and money). | Large public-sector IT organizations. | 2 companies |
| | Pino et al. (2010) | Reliability, construct, internal and external validity. | Small software organizations | 8 companies |
| | Amaral & Faria, 2010 | Content validity. | - | 1 company |
| | Kar et al., 2012 | Effectiveness and comfort | Small software organization. | 1 company |
| | Rapp et al., 2014 | Flexibility, efficiency, questions understandability, repeatability of results and meaningfulness of results. | - | 10 industrial projects and 27 projects in a semi-industrial environment. |
| | Kuvaja et al., 1999 | Participants feedback | - | - |
| | Abushama, 2016 | The use of the PAM-SMEs to guide process improvement with orientation to business objectives.<br>The applicability and suitability of the PAM-SMEs within SMEs. | ERP software packages, E-Learning software, E-Banking software | 3 companies |
| | Raza et al., 2012 | Reliability and validity | Open source projects | 2 projects |
| **Case study and Expert panel** | Burnstein et al. (1998) | - | - | 2 experts (3 three development groups) |
| **Expert panel** | MacMahon et al., 2015 | Utility, usability, scalability and generalizability, coverage of the requirements | - | 5 experts |
| **Comparison with other assessment method** | Karvonen et al., 2012 | Efficacy | - | - |
| **Not informed** | Göbel et al., 2013 | Functionality, usability, fit with the organization (the method is tailored for SME's within ITSM area), performance. | - | - |

**Table 12 Articles that report the evaluation methodology**